# Public Debate on Metallic Hydrogen to Boost High Pressure Research

Hua Y. Geng

National Key Laboratory of Shock Wave and Detonation Physics, Institute of Fluid Physics, CAEP; P.O. Box 919-102 Mianyang, Sichuan P. R. China, 621900

***Instead of praises from colleagues, the claim of observation of metallic hydrogen at 495 GPa by Dias and Silvera was met with skeptism, and grew into a public debate at the International Conference on High-Pressure Science and Technology, AIRAPT26. We briefly review this debate, and extend the topic to show that this disputation could be an oppurtunity to benefit the whole high pressure community.***

Debate is critical in modern society. It always goes on with consistency in logic and accuracy in facts, thus is a powerful tool to eliminate bias, extend knowledge boundary, and/or even expose falsity and fraud. Debate is a civilization approach for making decision nowadays. Through it, we can overcome our inherent limitations as an individual, and avoid from making serious mistakes. Public debate also played a pivotal role in history of science. One of the well-known cases includes the Bohr-Einstein debate, which greatly boosted the development of quantum mechanics. Now, a similar debate appeared at AIRAPT26, an important gathering of global high-pressure experts, on a recent claim of metallic hydrogen.

Metallic hydrogen (MH) refers to hydrogen in a metallic state [1]. To obtain it is dubbed as the *Holy Grail* in the high-pressure community. Quantum mechanics tells every material will become metallic under high enough compression. Therefore, metallization of hydrogen itself will not be much different from other elements at the very beginning. The thing became fascinating, however, when Ashcroft added a flavor into it and predicted that MH could be a room-temperature superconductor [2], as well as the accompanying bizarre and interesting phenomena when protons also become quantum [3,4]. These, together with the possibility that MH could be a strong explosive with ultra-high energy density, make MH a wonder material that attracts every experimentalist in the high-pressure community to pursue. Nonetheless, the challenge is tremendous, due to the notorious activity of hydrogen. Modern accurate theory put the transition into MH at a pressure about 500 GPa [5,6], whereas the most advanced DAC is limited to ~400 GPa on hydrogen. Thus it is a surprise to the high-pressure community when Dias and Silvera (DS) claimed they achieved a pressure as high as 495 GPa and obtained the MH in laboratory [7]. Their statement immediately caused backfire, and at least four leading groups in this field expressed disagreement [8-12]. Then, this fierce debate eventually went to public at the AIRAPT26 conference, held in Beijing this August.

The debate and dispute was mainly focused on three points: (1) the pressure calibration problem, (2) the credibility of the diamond Raman spectrum that gives the pressure of 495 GPa, and (3) whether reflectance alone is sufficient to assign MH. Without any internal pressure calibration, DS relied on a very special secondary pressure scale, the linear extrapolation of the load curve, to guide the DAC loading when in the blind stage (above 335 GPa) [7]. In principle, this is not a problem if the pressure of the final state can be reliably determined. But the possible interference from fictitious Raman peaks, which are frequently encountered at high pressures before the diamonds break, as Eremets pointed out [10], suggests that just one Raman spectrum is not enough to completely pin down the final pressure trustfully. People thus have to resort to the peculiar linear load curve to establish their judgment, which unfortunately has not been widely tested and accepted [8-11]. That is the reason why Loubeyre [11] and others want to see the continuity variation in the pressure scale and the pressure distribution across the chamber before accepting the claimed pressure record. This is understandable. Scarce data is always concomitant with unknown uncertainty, and is less convincing.

Fortunately, it is not a problem without solution. In AIRAPT26, an important proposal was announced to establish an international pressure standard to solve this problem. Therefore in the near future this kind of dispute on pressure calibration could be greatly reduced by bringing all different work onto the same level of pressure standard. This, of course, takes time. A realistic option available currently to resolve the disputation is to reproduce more reliable data from DS side. Reproducibility is the only answer to suspicion. It is challenging but not very difficult for DS, considering their reported high success ratio to reach a pressure above 400 GPa on hydrogen [9]. The reported Raman spectrum of diamond in [7] should be the first that needs to repeat [10]: it should present there as long as a pressure of 495 GPa is reached, no matter on hydrogen or on other inert materials. This spectrum can be verified or disproved by DS or by other groups. The concern about the load curve can also be eliminated by sharing their unique DAC device with the community or inviting a third party to participate in their experiments.

The basis for DS to claim the MH is the observed high reflectance. No one doubts they have observed the metal-like reflectance. The key problem here is what causes the high reflectivity. Eremets [10] and Loubeyre [11] have proposed their respective interpretations. Eremets also raised concerns about pressure measurements and noticed that he has observed reflection and semi-metallic behavior of hydrogen at lower pressures of 360 GPa and 100 K (arXiv:1601.04479 and arXiv:1708.05217) and it is strange for him that DS found a reflection only at 500 GPa and 80K. Nonetheless, DS insist it must come from MH by a Drude model analysis [7,12]. Fitting of the reflectance data to a Drude model could be problematic, since (*i*) MH at around 500 GPa does not behave like a free-electron metal at the low energy regime and thus cannot fit into a Drude model [13], and (*ii*) as Borinaga *et al.* pointed out, there is a large space of ambiguity in this nonlinear fitting with just two data points [13]. Nonetheless, the fact that the two lowest energy reflectance points match very well with a recent independent theoretical analysis [13], and another two points at higher energy, though without correction for the diamond absorption, qualitatively follow the predicted depression due to a unique interband plasmon at 6.2 eV in MH, is an encouraging message for DS. But a careful and reliable diamond absorption correction [8] must be made before one can tell whether it really corresponds to the MH fingerprint in reflectivity or not. In addition, hydrogen experiments often end with the development of incipient cracks in diamond anvils that lead to loss of hydrogen sample, and the Drude-like reflectance spectra could actually come from the metallic gasket filling the empty sample chamber. In particular, the IR spectra measured by Loubeyre *et al.* show different data than those reported by Silvera *et al.* in the pressure range of overlap [11], which further undermines the credibility of [7]. To demonstrate the presence of hydrogen sample and the reflectance indeed comes from hydrogen, DS need to show the diagnostic hydrogen Raman peaks during releasing pressure.

A clear message from this public debate is that we are now very near the discovery of MH. DS might have taken a leading position in this experimental race. But it is still too early for DS to make the final claim [8-11]. They must present a reliable pressure calibration, demonstrate the retainment of hydrogen sample, and reproduce the results. Substantial measurements other than the reflectance are also required.

One important aspect was ignored or downplayed in this debate—the metastability and recovery of MH at ambient conditions. Nellis raised the importance of this topic, but met with weak resonance. In [7], DS referred to [14] for the justification of recoverability of MH. This could be wildly optimistic. Actually, an exploration of the possible energy barriers in MH at ambient pressure with accurate modern density functional theory (DFT) and NEB method unfortunately revealed that MH should be highly unstable at ambient pressure [15]. We extend a similar analysis at relatively high pressures here, for a purpose to investigate at what pressure MH can be metastable. Both the degenerate Cs-IV and Fddd phases have been studied. The conclusion is the same, and thus we only focus on Fddd below.

At first we explored the superheating limit of MH down to 200 GPa, using AIMD in PBE approximation of DFT as implemented in VASP. With a large cell containing 480H, a *k*-point grid of 2×2×2, and an energy cutoff of 600 eV, we found that the classical superheating temperature of MH within this pressure range is very low, as shown in the left panel of Fig. 1. Note that if included the nuclear quantum effects of protons, the superheating limit should lower further. The indication is that one cannot have MH, even in a metastable state, at a pressure less than 200 GPa and at a temperature as low as 100 K. This conclusion is further strengthened by an NEB energy barrier calculation using both PBE and vdW-DF functional. It is well known that DFT has some problems in describing the $H_2$ dissociation. But both accurate QMC calculations and dynamic compression experiment showed that the true physics in dense hydrogen around dissociation should be well bracketed by PBE and vdW-DF functional [16]. We thus employed both methods to avoid possible bias. The results for 315 GPa is given as the right panel in Fig. 1. A very weak barrier (in both PBE and vdW-DF) of about 0.03 eV/H is observed. This value corresponds to a temperature of 348 K, being consistent with the superheating temperature. Nonetheless, this barrier reduces rapidly with further decreasing pressure. And the conclusion is MH cannot be recovered to low pressure with traditional methods.

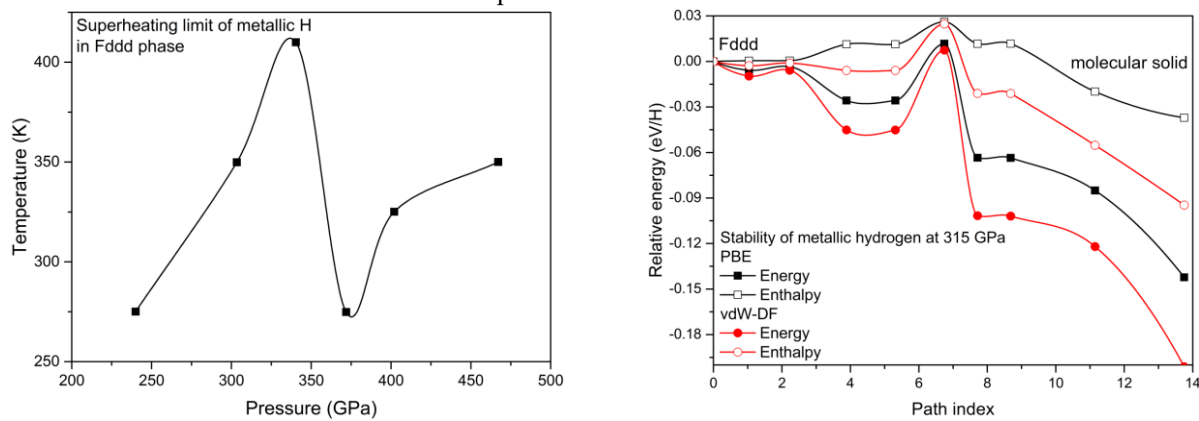

Figure 1. Classical superheating limit (left) and energy barrier (right) calculated for MH in Fddd phase using AIMD and NEB method at the DFT level.

As mentioned above, the importance of MH is not in the metallization itself, but mainly in the potential application of MH as a test model for quantum many-body theory at very high density, and as a room temperature superconductor or a high energy density material, as well as the capability to turn this wonder material into real productivity. It might have a huge and deep impact on the future of mankind. There is no precedent in high pressure community with such a possible direct entanglement in civilization development and to drive the society forward. To obtain MH at high pressure condition is challenging enough, and to retrieve it back to ambient conditions is even much more challenging, requiring unconventional and extraordinary creativity. It is hard to say of having grabbed the *Holy Grail* by just observing it. The public debate on MH, fortunately, could greatly boost high pressure research by these top experimentalists through unveiling their *secret weapons* and special techniques, as well as sharing their unique experience of achieving such high pressures on such a difficult material. This still ongoing public debate undoubtedly will attract and gather talent young scientists continuously flowing into this promising field, to foster and create novel techniques that will eventually pave the way towards the ultimate goal.

ACKNOWLEDGMENTS

H.Y.G. acknowledges support from the National Natural Science Foundation of China under Grant No. 11672274 and the NSAF under Grant No. U1730248.